\documentclass[12pt]{cernrep}
\usepackage{graphicx}
\usepackage{floatflt}
\usepackage{color}
\usepackage{colortbl}

\newcommand{\Pt}{{P_t}}

\newcommand{\dphi}{\Delta\phi}
\newcommand{\phigj}{\phi_{(\gamma,~jet)}}

\newcommand{\Ptgj}{$\Pt^{\gamma}$ and $\Pt^{Jet}$~~}

\newcommand{\la}{\langle}
\newcommand{\ra}{\rangle}
\newcommand{\gpj}{~"$\gamma+Jet$"~}

\newcommand{\rrr}{\to} %{\rightarrow}

\newcommand{\pth}{\hat{p}_{\perp}^{\;min}}
\newcommand{\Db}{\Pt(O+\eta>5.0)}
\newcommand{\Ptg}{\Pt^{\gamma}}

\newcommand{\Fptgj}{(\Pt^{\gamma}-\Pt^{Jet})/\Pt^{\gamma}}

\newcommand{\coltab}{0.69}

\newcommand{\ptgj}{$~\Pt^{\gamma} - \Pt^{Jet}~$}

\newcommand{\lt}{\!<\!}
\newcommand{\gt}{\!>\!}

\newcommand{\hmm}{\hspace*{-1.3mm}}
\newcommand{\han}{\hspace*{.58cm}}
\newcommand{\has}{\hspace*{.52cm}}
\newcommand{\hass}{\hspace*{.4cm}}
\newcommand{\had}{\hspace*{.54cm}}

\newcommand{\hcn}{\hspace*{.57cm}}

\newcommand{\hcd}{\hspace*{.52cm}}

\newcommand{\Gvc}{\footnotesize{$(GeV/c)$} }

\suppressfloats[!]

\def\baselinestretch{1.0}

\begin{document}
%==========================================================================
%      
\noindent                                                                   
\large{\bf \gpj process application for setting the absolute scale of jet energy
and determining the gluon distribution at the LHC.} \\[10pt]
%                                                                         |
%==========================================================================
\normalsize
{\hspace*{3cm} D.~Bandurin, V.~Konoplyanikov, N.~Skachkov}\\
{\it ~\hspace*{3cm} Joint Institute for Nuclear Research, Dubna, Russia}
%\maketitle
\vskip2mm
\normalsize
\hspace*{5cm}{\bf Abstract}\\[5pt]
\noindent
The possibility of jet energy scale setting 
at the CMS calorimeter by using \gpj process is studied.
The estimation of the number of \gpj events suitable for determination of
gluon distribution inside a proton in a new kinematic region of $x$, $Q^2$
variables beyond the one covered by HERA data is also presented.

\thispagestyle{plain}

\section{Introduction.}

Basing on the selection criteria introduced for the first time in \cite{1}--\cite{BKS_P1}
(see below Section 2), the background events suppression factors, 
signal events selection efficiencies and the number of the events, that can be collected
at integrated luminosity $L_{int}=3~fb^{-1}$ are determined here.

It is also shown that \gpj events, being collected at LHC,
would provide us with the data sufficient
for an extraction of gluon distribution function in a proton.
A new region of $2\cdot10^{-4}\leq x \leq 1$ and
$1.6\cdot 10^3\leq Q^2\leq8\cdot 10^4 ~(GeV/c)^2$ can be covered.
The rates of $g\,c\to \gamma^{dir} \,+\,Jet$ events are also given.

%-------------------------------------------------------------------
%                                                                  |
\section{Definition of selection cuts for physical variables and
the scalar form of the $\Pt$ balance equation.}
                                                                  %|
%           3.2                                                    |
%----------------------------------------------------------------

\noindent
1. We shall select the events with one jet and one ``$\gamma^{dir}$-candidate''
(in what follows we designate it as $\gamma$ and call the
``photon'' for brevity and only in Section 3, devoted to the backgrounds, we denote
$\gamma^{dir}$-candidate by $\tilde{\gamma}$) with\\[-12pt]
\begin{equation}
\Pt^{\gamma} \geq 40~ GeV/c~\quad {\rm and} \quad \Pt^{Jet}\geq 30 \;GeV/c.
\label{eq:sc1}
\end{equation}
%\noindent
The electromagnetic calorimeter (ECAL) signal can be considered as a candidate for a direct photon
if it fits inside the 5$\times$5 ECAL crystal cell window having a cell
with the highest $\Pt$ $\gamma/e$ in the center (\cite{CMS_EC}).

The jet is defined here according to the PYTHIA \cite{PYT}
jetfinding algorithm LUCELL.
The jet cone radius R in the $\eta-\phi$ space counted from the jet initiator cell (ic) is
taken to be $R_{ic}=((\Delta\eta)^2+(\Delta\phi)^2)^{1/2}=0.7$.
%Below we shall also use the jet radius counted
%from the center of gravity (gc) of the jet, i.e. $R_{gc}$.

\noindent
2. To suppress the background processes, i.e. to select mostly the events with ``isolated''
photons and to discard the events with fake ``photons'' (that
may originate as ``$\gamma^{dir}$-candidates'' from meson decays, for instance), we restrict

a) the value of the scalar sum of $\Pt$ of hadrons and other particles surrounding
a ``photon'' within a cone of $R^{\gamma}_{isol}=( (\Delta\eta)^2+(\Delta\phi)^2)^{1/2}=0.7$
(``absolute isolation cut")\\[-14pt]
\begin{equation}
\sum\limits_{i \in R} \Pt^i \equiv \Pt^{isol} \leq \Pt_{CUT}^{isol};
\label{eq:sc2}
\end{equation}
\vspace{-2.6mm}

b) the value of a fraction (``relative isolation cut'')\\[-12pt]
\begin{equation}
\sum\limits_{i \in R} \Pt^i/\Pt^{\gamma} \equiv \epsilon^{\gamma} \leq
\epsilon^{\gamma}_{CUT}.
\label{eq:sc3}
\end{equation}

\noindent
3. To be consistent with the application condition of the NLO
formulae, one should avoid an infrared dangerous region and take care of
$\Pt$ population in the region close to a $\gamma^{dir}$-candidate we also restrict
in accordance with \cite{Fri} and \cite{Cat}  the scalar sum of $\Pt$ of particles
 around a ``photon'' within a cone of a smaller radius $R^{\gamma}_{singl} = 0.175 = 1/4
\,R_{isol}^{\gamma}$:\\[-4mm]
\begin{equation}
\sum\limits_{i \in R^{\gamma}_{singl}} \Pt^i \equiv \Pt^{singl} \leq 2~ GeV/c
~~~~~(i\neq ~\gamma-dir).
\label{eq:sc4}
\end{equation}

\noindent
4. We accept only the events having no charged tracks (particles)
with $\Pt>1~GeV/c$ within the $R^{\gamma}_{singl}$ cone around the $\gamma^{dir}$-candidate.

\noindent
5. %We also consider the structure of every event with the photon
%candidate at a more precise level of the 5$\times$5 crystal cell window (size of one D0
%HCAL tower) with a cell size of 0.0175$\times$0.0175. 
To suppress the background events with photons resulting from high-energy $\pi^0$, $\eta$, $\omega$
and $K_S^0$ meson decays, we require the absence of a high $\Pt$ hadron
in the calorimeter tower containing the $\gamma^{dir}$-candidate:\\[-10pt]
\begin{equation}
\Pt^{hadr} \leq 5~ GeV/c.
\label{eq:sc5}
\end{equation}

\noindent
At the PYTHIA level of simulation this cut may effectively take into account 
the imposing of an upper cut on the hadronic calorimeter (HCAL) signal in the towers behind
the ECAL tower fired by the direct photon.
%We can not reduce this value down to, for example, 2-3 $GeV/c$, because
%a hadron with $\Pt$ below 2-3 $GeV/c$ deposits with high probability most of its energy in ECAL and
%may not reveal itself in HCAL. The value 5 $GeV/c$ is chosen with account of possible
%loss of hadron energy in ECAL (see Fig.~10A of Appendix 6).

\noindent   % text are changed (S)
6. We select the events with the vector $\vec{\Pt}^{Jet}$ being ``back-to-back" to
the vector $\vec{\Pt}^{\gamma}$ (in the plane transverse to the beam line)
within $\dphi$ defined by the equation:\\[-12pt]
\begin{equation}
\phigj=180^\circ \pm \Delta\phi \quad (\Delta\phi =15^\circ, 10^\circ, 5^\circ)
\label{eq:sc7}
\end{equation}
($5^\circ$ is one HCAL tower  size in $\phi$), where $\phigj$ is the angle
between the \Ptgj vectors: 
$\vec{\Pt}^{\gamma}\vec{\Pt}^{Jet}=\Pt^{\gamma}\Pt^{Jet}\cdot cos(\phigj)$,
where ~$\Pt^{\gamma}=|\vec{\Pt}^{\gamma}|,~~\Pt^{Jet}=|\vec{\Pt}^{Jet}|$.

\noindent
7. The initial and final state radiations (ISR and FSR) manifest themselves most clearly
as some final state mini-jets or clusters activity.
To suppress it, we impose a new cut condition that was not formulated in
an evident form in previous experiments: we choose the \gpj events
that do not have any other
jet-like or cluster high $\Pt$ activity  by selecting the events with the values of
$\Pt^{clust}$ (the cluster cone $R_{clust}(\eta,\phi)=0.7$), being lower than some threshold
$\Pt^{clust}_{CUT}$ value, i.e. we select the events with\\[-10pt]
\begin{equation}
\Pt^{clust} \leq \Pt^{clust}_{CUT}
\label{eq:sc8}
\end{equation}
($\Pt^{clust}_{CUT}=15, 10, 5 ~GeV/c$ are most efficient as will be shown in Section 3).
Here, the clusters are found by one and the same jetfinder LUCELL.

Now we pass to another new quantity (introduced also for the first time in \cite{1}--\cite{BKS_P1}) 
that can be measured at the experiment.

\noindent
8. We limit the value of the modulus of the vector sum of $\vec{\Pt}$ of all
particles, except those of the \gpj system, that fit into the region $|\eta|\lt5.0$ covered by
the ECAL and HCAL, i.e., we limit the signal in the cells ``beyond the jet and photon'' region
by the following cut:\\[-10pt]
\begin{equation}
\left|\sum_{i\not\in Jet,\gamma-dir}\!\!\!\vec{\Pt}^i\right| \equiv \Pt^{out} \leq \Pt^{out}_{CUT},
~~|\eta_i|\lt5.0.
\label{eq:sc9}
\end{equation}
%\stackrel{def}{=}

\noindent
The importance of $\Pt^{out}_{CUT}$ and $\Pt^{clust}_{CUT}$
for selection of events with a good balance of \Ptgj and for
the background reduction will be demonstrated in Section 3.

Below the set of selection cuts 1 -- 8 will be referred to as
``Selection 1''. The last two of them, 7 and 8, are new criteria \cite{1}--\cite{BKS_P1}
not used in previous experiments. In addition to them one more new object, introduced in 
\cite{BKS_P1} and named an ``isolated jet'',  will be discussed.

\noindent
9. We also involve a new requirement of ``jet isolation'',
i.e. the presence of a ``clean enough'' (in the sense of limited $\Pt$
activity) region inside the ring (of $\Delta R=0.3$ or of approximately a size of 
three calorimeter towers) around the jet.  
Following this picture, we restrict the ratio of the scalar sum
of transverse momenta of particles belonging to this ring, i.e.\\[-5pt]
\begin{equation}
\Pt^{ring}/\Pt^{Jet} \equiv \epsilon^{jet}, \quad {\rm where ~~~~ }
\Pt^{ring}=\sum\limits_{\footnotesize i \in 0.7\lt R \lt1} |\vec{\Pt}^i|.
\label{eq:sc10}
\end{equation}
~\\[-4pt]
($\epsilon^{jet}\leq 3-5\%$).
The set of events that %satisfy this restriction
pass cuts 1 -- 9 will be called ``Selection 2''.

The exact values of the cut parameters $\Pt^{isol}_{CUT}$,
$\epsilon^{\gamma}_{CUT}$, $\epsilon^{jet}$, $\Pt^{clust}_{CUT}$, $\Pt^{out}_{CUT}$
 will be specified below, since they may be
different, for instance, for various $\Pt^{\gamma}$ intervals
(being looser for higher  $\Pt^{\gamma}$).

\noindent 
10. One can expect reasonable results of the jet energy calibration procedure
modeling and subsequent practical realization
 only if one uses a set of selected events with small $\Pt^{miss}$ caused by neutrinos 
instrumental/material features of the detector. So, we also use
the following cut:\\[-22pt]
\begin{eqnarray}
\Pt^{miss}~\leq \Pt^{miss}_{CUT}.
\label{eq:sc11}
\end{eqnarray}
The aim of the event selection with small $\Pt^{Jet}_{(\nu)}$
is quite obvious: we need a set of events with a reduced
$\Pt^{Jet}$ uncertainty due to possible presence of a non-detectable
neutrino contribution to a jet, for example \cite{BKS_P1}.

To conclude this section, let us rewrite
the scalar $\Pt$ balance equation from \cite{BKS_P1} 
with the notations introduced there in the form
more suitable to present the final results:\\[-18pt]
%For this purpose we shall write equation (16) in the following scalar form: \\[-15pt]
\begin{eqnarray}
\frac{\Pt^{\gamma}-\Pt^{Jet}}{\Pt^{\gamma}}=(1-cos\dphi) %\phi_{(\gamma,jet)}
+ \Db/\Pt^{\gamma}, \label{eq:sc12}
\label{eq:sc_bal}
\end{eqnarray}
\vskip-10pt
\noindent
where
$\Db\equiv (\vec{\Pt}^{O}+\vec{\Pt}^{|\eta|>5.0)})\cdot \vec{n}^{Jet}$ ~~ with ~
$\vec{n}^{Jet}=\vec{\Pt}^{Jet}/\Pt^{Jet}$. Here
$\Pt^O$ is a total transverse momentum of all particles beyond \gpj system in the 
$|\eta|\lt5.0$ region and $\Pt^{|\eta|>5.0}$ is a total transverse momentum of all
particles flying in the direction of a non-instrumented forward
part ($|\eta|>5.0$) of the D0 detector.
%(28)

As shown in Section \cite{BKS_P1}, the first term on the
right-hand side of equation (\ref{eq:sc_bal}), i.e. $(1-cos\dphi)$ is negligibly
small and tends to decrease fast with  growing $\Pt^{Jet}$. 
So, the main contribution to the $\Pt$ disbalance in the
\gpj system is caused by the term $\Db/\Pt^{\gamma}$.

%----------------------------------------------------------------------
%                                                                      |
\section{Detailed study of background suppression.}                   %|
%    Chapter 8.                                                        |
%-----------------------------------------------------------------------

To estimate the background for the signal events, we carried out the simulation
with a mixture of all QCD and SM subprocesses with large
cross sections existing in PYTHIA
%%%
\footnote{ PYTHIA~5.7 version with default CTEQ2L parametrisation
of structure functions is used here.}, 
%%%
namely, ISUB=1, 2, 11--20, 28--31, 53, 68, which 
can lead to a large background for our main ``signal'' subprocesses (\ref{eq:1a}) and (\ref{eq:1b})
(ISUB=14 and 29 in PYTHIA)
%%%
\footnote{A contribution of another possible NLO channel $gg\rrr g\gamma$
(ISUB=115 in PYTHIA) was found to be still negligible even at LHC energies.}:
%%%
\\[-5mm]
``Compton-like'' process\\[-8mm]
\begin{eqnarray}
\hspace*{1.04cm} qg\to q+\gamma %\hspace*{7.5cm} 
\label{eq:1a}
\end{eqnarray}
\vspace{-3mm}
and the ``annihilation'' process\\[-15pt]
\begin{eqnarray}
\hspace*{1.02cm} q\overline{q}\to g+\gamma.  %\hspace*{7.4cm} 
\label{eq:1b}
\vspace{-3mm}
\end{eqnarray}

Three generations  with the abovementioned set of subprocesses were performed, each with
different minimal values of $\Pt$ appearing in the final state
of the hard $2\to 2$ subprocess, i.e $\pth = CKIN(3)$ parameter in
PYTHIA that practically coincides with $\Ptg$ in the case of
signal direct photons production (compare lines 2 and 3 from the column ``$S$'' of Table 2). 
These values were $\pth$ =40 $GeV/c$, 100 and 200 $GeV/c$. By 40 million events were
generated for three $\pth$ values respectively. The cross sections
of the abovementioned subprocesses define the rates of corresponding physical
events and, thus, appear here as weight factors.

%\noindent
We selected ``$\gamma^{dir}$-candidate +1 Jet'' events 
with $\Pt^{Jet}\gt30$ $GeV/c$ containing one $\gamma^{dir}$-candidate
(denoted as ${\tilde{\gamma}}$) to be identified by the detector as an isolated photon 
%%%
\footnote{For brevity 
we denote the direct photon and the ``$\gamma^{dir}$-candidate'' by  the same
symbol ``$\tilde{\gamma}$''.}
%%%
with $R^{\gamma}_{isol}=0.7$ and $\Pt^{\tilde{\gamma}}\geq 40$ (~100 and ~200) $~GeV/c$
for the generation with $\pth = 40$ (~100 and ~200) $~GeV/c$ respectively
(see below cut 3 $\Pt^{\tilde{\gamma}}\geq\pth$ of Table \ref{tab:sb0}.).
Here and below, speaking about the $\gamma^{dir}$-candidate, 
we actually mean a signal that may be registered in the
$5\times 5$ ECAL crystal cell window having the cell with the highest $\Pt$ photon or electron
($\gamma/e$) in its center.
All these photon candidates were supposed to satisfy isolation criteria of \cite{BKS_P1}
with the values given in Table \ref{tab:sb0}:
$\Pt^{isol}_{CUT} = 2 ~GeV/c$ and $\epsilon^{\tilde{\gamma}}_{CUT}=5\% $.

We apply the cuts from Table~\ref{tab:sb0} one after another on the observable
physical variables. The influence of these cuts on the signal-to-background
 ratio $S/B$ is presented in Tables 2, 5--7.
%for the case of $\Pt^{\tilde{\gamma}}\geq 100 ~GeV/c$.

Tables \ref{tab:sb4} and \ref{tab:sb1} are complementary to each other.
The numbers in the left-hand column (``Cut'') of Table~\ref{tab:sb4},
coincide with the numbers of cuts listed in Table~\ref{tab:sb0}.

%\noindent
The second and third columns contain respectively the numbers of 
signal direct photons ($S$)
\footnote{Their number coincide starting from line 3 of Table 1 with the number of events
with (\ref{eq:1a}) and (\ref{eq:1b}) fundamental $2\to2$ subprocesses of direct photon production.}
 and background $\gamma^{dir}-$candidates ($B$)
left in the sample of events after application of each cut. The numbers of
background  events $B$ do not include events with electrons. 
%which contribution will be discussed later. 
Their numbers in the samples are presented separately in the last right-hand column ``$e^\pm$''.
The other columns of Table \ref{tab:sb4} include efficiencies
$Eff_{S(B)}$ (with their errors) defined as a ratio
of the number of signal (background) events that passed under a cut
(1--17) to the number of the preselected events (1st cut of this table).
They are followed by the column containing the values of $S/B$
(without account of events with electrons that fake  direct photons).
~\\[-10mm]
\begin{table}[h]
%\nonumber
\small
\caption{List of the applied cuts used in Tables \ref{tab:sb4}, 5--7.}
\begin{tabular}{lc} \hline
\label{tab:sb0}
{\bf 0}. No cuts; \\
{\bf 1}. $a)~\Pt^{\tilde{\gamma}}\geq 40 ~GeV/c, b)~|\eta^{\tilde{\gamma}}|\leq 2.61,
c)~\Pt^{jet}\geq 30 ~GeV/c, d)~\Pt^{hadr}\!<5 ~GeV/c^{\;\ast}$;\\
%\hspace*{3mm} \\
{\bf 2}. $\epsilon^{\tilde{\gamma}} \leq 15\%$;
\hspace*{1.87cm} {\bf 11}. $\Pt^{clust}<20 ~GeV/c$; \\
{\bf 3}. $\Pt^{\tilde{\gamma}}\geq\pth$;
\hspace*{1.5cm} {\bf 12}. $\Pt^{clust}<15 ~GeV/c$; \\
%{\bf 4}. $\Pt^{sum}\!<1~ GeV/c^{\;\ast\ast}$ ;
{\bf 4}. $\epsilon^{\tilde{\gamma}} \leq 5\%$;
\hspace*{2.03cm} {\bf 13}. $\Pt^{clust}<10 ~GeV/c$; \\
{\bf 5}. $\Pt^{isol}\!\leq 2~ GeV/c$;
\hspace*{0.67cm} {\bf 14}. $\Pt^{out}<20 ~GeV/c$; \\
{\bf 6}. $Njet\leq3$;
\hspace*{1.83cm}  {\bf 15}. $\Pt^{out}<15 ~GeV/c$; \\
{\bf 7}. $Njet\leq2$;
\hspace*{1.85cm}  {\bf 16}. $\Pt^{out}<10 ~GeV/c$;\\
{\bf 8}. $Njet=1$;
\hspace*{1.84cm} {\bf 17}. $\epsilon^{jet} \leq 5\%$.\\
{\bf 9}. $\dphi<15^\circ$; \\
%\hspace*{1.9cm} {\bf 19}. $\epsilon^{jet} \leq 2\%$.\\
{\bf 10}. $\Pt^{miss}\!\leq10~GeV/c$;\\\hline
\footnotesize{${\;\ast}~\Pt$ of a hadron in the 5x5 ECAL cell window containing 
the $\gamma^{dir}$-candidate in the center.}\\
% \footnotesize{${\;\ast\ast}$ A scalar sum of $\Pt$ in the 5x5 ECAL cell window 
% in the region out of a smaller 3x3 window containing $\tilde{\gamma}$.}\\[-3mm]
\end{tabular}
\vskip-2mm
\end{table}

\normalsize
From the first line of Table 14 we see that without imposing any cut
the number of background events $B$ (the 3rd column)  
exceeds the number of signal events $S$ (the 2nd column) by 5 orders of magnitude.
The relative isolation cut 2 ($\epsilon^{\tilde{\gamma}} \leq 15\%$) makes the 
$S/B$ ratio equal to 0.28.
Cut 3 ($\Pt^{\tilde{\gamma}}\geq\pth$) improves the $S/B$ ratio to 0.71.
Relative isolation cut 4 and then the absolute isolation cut 5 make
the $S/B$ ratio to be equal to 1.50 and 1.93, respectively. The requirement
of only one jet being present in the event (cut 8) results in the value $S/B=5.96$.
The ratio $S/B$ is increased by the cut $\dphi<15^\circ$ to 6.54
(cut 9) and at the same time the number of signal events is decreased only by $5\%$.
%This is in agreement with the phenomenon of the concentration of events in
%the small $\dphi$ angle at large $\Pt^{\tilde{\gamma}}$ already mentioned in Section 5.

\begin{figure}[htbp]
\vspace{-1.5cm}
\hspace{-9mm} \includegraphics[width=16cm,height=7.5cm]{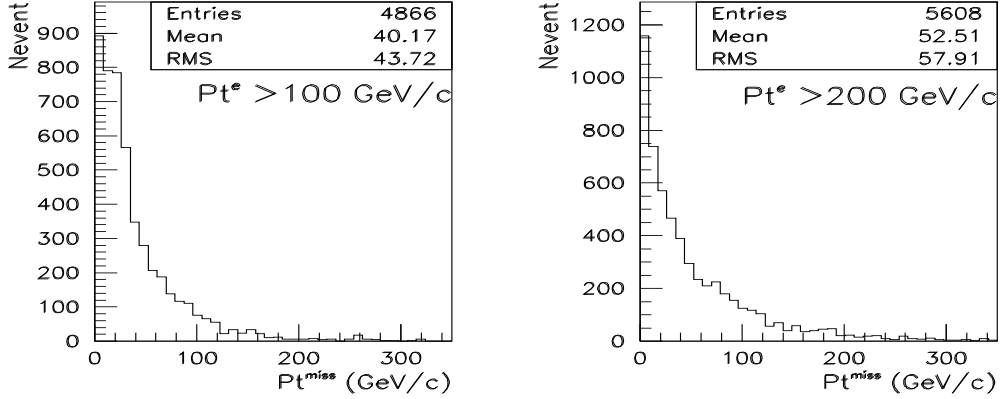}
\vspace{-1.5cm}
\caption{\hspace*{0.0cm} Distribution of events over $\Pt^{miss}$
in  events with energetic $e^\pm$`s appearing as direct photon candidates for 
the cases ~$\Pt^e\geq100~ GeV/c~$
and $~\Pt^e\geq200~ GeV/c$~ (here are used events satisfying cuts 1--3 of
Table \ref{tab:sb0}).}
\label{fig:ptmiss} %Fig. 20
\vskip-5mm
\end{figure}

In line 10 we used the $\Pt^{miss}_{CUT}$ cut, described in Section 2,
to reduce uncertainty of $\Pt^{Jet}$ due to a possible neutrino
contribution to a jet, for example. It also reduces the contribution to background
%Here it is applied against the processes based (at the parton level) on the
from the decay subprocesses ~$q\,g \to q' + W^{\pm}$~ and ~$q\bar{~q'} \to g + W^{\pm}$~ 
with the subsequent decay $W^{\pm} \to e^{\pm}\nu$ that leads to a substantial $\Pt^{miss}$ value.
It is clear from the distributions over $\Pt^{miss}$ for two
$\Pt^e$ intervals presented in Fig.~\ref{fig:ptmiss}. From the last column ($e^\pm$) of 
Table \ref{tab:sb4} one can see that $\Pt^{miss}_{CUT}$ cut
(see line 10) reduces strongly (5 times) the number of events containing $e^\pm$ as
direct photon candidates. So, $\Pt^{miss}_{CUT}$ would make a noticeable improvement
of the total $S/B$ ratio. 
\\[-2mm]

The cuts 11--13 show step-by-step influence of $\Pt^{clust}_{CUT}$.
The reduction of $\Pt^{clust}_{CUT}$ to $10 ~GeV/c$ (cut 13)
results in significant improvement (about 3 times as compared with line 10)
of the $S/B$ ratio to 17.64. Further
reduction of $\Pt^{out}_{CUT}$ to $10 ~GeV/c$ (cut 16) improves $S/B$ to
22.67. The jet isolation requirement
 $\epsilon^{jet}<5\%$ (line 17) finally gives $S/B=24.46$
\footnote{Stricter isolation requirement $\epsilon^{jet}<2\%$ considered in \cite{BKS_P1}
would lead to $S/B = 31.1$.}.
The summary of Table \ref{tab:sb4} is presented in the middle section ($\pth=100 ~GeV/c$)
of Table \ref{tab:sb1} where line ``Preselected''
corresponds to the cut 1 of Table  \ref{tab:sb0} and correspondingly to the line
number 1 of  Table  \ref{tab:sb4} presented above.
The line ``After cuts'' corresponds to the line 16 of  Table \ref{tab:sb4} and 
line ``+jet isolation'' corresponds to the line 17.

Tables \ref{tab:bg_or_gr} and \ref{tab:bg_or_ms} show  the relative
contributions of four main (having the largest cross sections) fundamental QCD subprocesses 
$qg\to qg$, $qq\to qq$, $gg\to q\bar{q}$ and $gg\to gg$ into production of the background 
``$\gamma\!-\!brem$'' and ``$\gamma\!-\!mes$''  events selected 
by criteria 1--13 of Table 1 for three $\Ptg$ intervals.
%
%\footnote{Because of a shortage of statistics, we have not taken into consideration $\Pt^{out}_{CUT}$
%in Table \ref{tab:bg_or_gr}. It would be also desirable to perform analogous estimations for all
%types of the mesons (see Table 17) what is also the matter of statistics.}.
%. 
In some lines of Tables \ref{tab:bg_or_gr} and \ref{tab:bg_or_ms} the sum over contributions
from the four considered QCD subprocesses is less than 100$\%$. The remained percentages correspond to
other subprocesses (like $q\bar{q}\to q\bar{q}$).

It is useful to note from Tables \ref{tab:bg_or_gr} and \ref{tab:bg_or_ms}
that most of background events ($85\%$ at least) originate from  
$qg\to qg$ and $qq\to qq$ scatterings 
%with dominant contribution from the first process.
with an increase of contribution from the last one with growing $\Ptg$.
\def\baselinestretch{0.97}
\begin{table}[h]
\small
\begin{center}
\vskip-0.2cm
\caption{Values of significance and  efficiencies for $\hat{p}_{\perp}^{\;min}$=100 $GeV/c$.}
\vskip1mm
\begin{tabular}{||c||c|c|c|c|c|c||}                  \hline \hline
\label{tab:sb4}
Cut& $S$ & $B$ & $Eff_S(\%)$ & $Eff_{B}(\%)$  & $S/B$& $e^\pm$ \\\hline \hline
\rowcolor[gray]{\coltab}%
 0 & 19420 & 5356.E+6 &             &                 &0.00 &3.9E+6  \\\hline % 1
\rowcolor[gray]{\coltab}%
%Preselected
 1 & 19359 &  1151425 & 100.00 $\pm$ 0.00& 100.000 $\pm$ 0.000  &0.02 &47061\\\hline % 4
%epsilon<15%
 2 & 18236 &   65839  & 94.20 $\pm$ 0.97 &   5.718 $\pm$ 0.023  &0.28 &8809 \\\hline % 6
% Pt_gam > Pt_min
 3 & 15197 &    22437 &  78.50 $\pm$ 0.85&   1.949 $\pm$ 0.013&  0.71 &2507 \\\hline % 8
% v 5x5 vne 3x3
% 4 & 15174 &    19005 &  78.38 $\pm$ 0.85&   1.651 $\pm$ 0.012&  0.80 &2486\\\hline % 10
% gamma isol.
 4 & 14140 &     9433 &  73.04 $\pm$ 0.81&   0.819 $\pm$ 0.008&  1.50 &2210 \\\hline % 12
 5 &  8892 &     4618 &  45.93 $\pm$ 0.59&   0.401 $\pm$ 0.006&  1.93 &1331 \\\hline % 13
% Njet
 6 & 8572  &    3748  &  44.28 $\pm$ 0.57&   0.326$\pm$  0.005&  2.29 &1174  \\\hline % 14
 7 & 7663  &    2488  &  39.58 $\pm$ 0.53&   0.216$\pm$  0.004&  3.08 & 921  \\\hline % 15
 8 & 4844  &     813  &  25.02 $\pm$ 0.40&   0.071$\pm$  0.002&  5.96 & 505  \\\hline % 16
% dphi
 9 & 4634  &     709  &  23.94 $\pm$ 0.39&   0.062 $\pm$ 0.002&  6.54 & 406 \\\hline % 18
% ptmiss
10 &  4244 &     650  &  21.92 $\pm$ 0.37&   0.056 $\pm$ 0.002&  6.53 &  87 \\\hline % 20
% clu
11 &  3261 &      345 &  16.84 $\pm$ 0.32&   0.030 $\pm$ 0.002&  9.45 &53 \\\hline % 21
12 &  2558 &      194 &  13.21 $\pm$ 0.28&   0.017 $\pm$ 0.001& 13.19 &41 \\\hline % 22
13 &  1605 &       91 &   8.29 $\pm$ 0.22&   0.008 $\pm$ 0.001& 17.64 &26 \\\hline % 23
% out
14 &  1568 &       86 &   8.10 $\pm$ 0.21&   0.007 $\pm$ 0.001& 18.23 &26 \\\hline % 25
15 &  1477 &       77 &   7.63 $\pm$ 0.21&   0.007 $\pm$ 0.001& 19.18 &25 \\\hline % 26
\rowcolor[gray]{\coltab}%
16 &  1179 &       52 &   6.09 $\pm$ 0.18&   0.005 $\pm$ 0.001& 22.67 &22 \\\hline % 27
% jet isol.
\rowcolor[gray]{\coltab}%
17 &  1125 &       46 &   5.81 $\pm$ 0.18&   0.004 $\pm$ 0.001& 24.46 &21 \\\hline % 30
%\rowcolor[gray]{\coltab}%
%19 &   683 &       22 &   3.53 $\pm$ 0.14&   0.002 $\pm$ 0.000& 31.05 &13 \\\hline % 31
\hline \hline
\end{tabular}
\end{center}
\vskip-3mm
\noindent
\hspace*{5mm} \footnotesize{${\;\ast}$ The background ($B$) is considered here with no
account of contribution from the ``$e^\pm$ events'' in which $e^\pm$`s appear as 
$\gamma^{dir}$-candidates.}
\vskip-3mm
\end{table}

The simulation in PYTHIA also predicts that practically in all selected ``$\gamma\!-\!brem$'' events
``bremsstrahlung photons'' are produced in the final state of the fundamental subprocess.
Namely, they are radiated from the outgoing quarks in the case of the first three subprocesses
and appear as the result of string breaking in the case of $gg\to gg$ scattering which,
naturally, gives a small contribution into \gpj events production.
~\\[-5mm]
\begin{table}[h]
\begin{center}
\vskip-3mm
\caption{Relative contribution (in per cents) of different QCD subprocesses into
the ``$\gamma\!-\!brem$'' events production.}
\normalsize
\vskip.1cm
\begin{tabular}{|c||c|c|c|c|}                  \hline \hline
\label{tab:bg_or_gr}
$\Ptg$& \multicolumn{4}{c|}{fundamental QCD subprocess} \\\cline{2-5}
 \Gvc & $qg\to qg$ & $qq\to qq$ & $gg\to q\bar{q}$& $gg\to gg$  
\\\hline \hline
 40--71   &  70.6$\pm$ 8.7 & 21.1$\pm$ 3.8 &  5.1$\pm$ 1.6 &  2.6$\pm$ 1.0 \\\hline %99.4 total 
 71--141  &  67.5$\pm$ 7.3 & 23.6$\pm$ 3.5 &  4.2$\pm$ 1.2 &  2.6$\pm$ 0.9  \\\hline %97.9 
141--283  &  58.7$\pm$ 9.0 & 30.7$\pm$ 5.7 &  1.8$\pm$ 1.0 &   ---   \\\hline\hline  %91.2
\end{tabular}
\end{center}
\vskip-4mm
\end{table}

\begin{table}[h]
\begin{center}
\vskip-2mm
\caption{Relative contribution (in per cents) of different QCD subprocesses into
the ``$\gamma\!-\!mes$'' events production.}
\normalsize
\vskip.1cm
\begin{tabular}{|c||c|c|c|c|}                  \hline \hline
\label{tab:bg_or_ms}
$\Ptg$& \multicolumn{4}{c|}{fundamental QCD subprocess} \\\cline{2-5}
 \Gvc & $qg\to qg$ & $qq\to qq$ & $gg\to q\bar{q}$& $gg\to gg$ 
\\\hline \hline
 40--71   &  65.2$\pm$ 9.9 & 20.1$\pm$ 4.5 &  7.1$\pm$ 2.5 &  7.2$\pm$ 2.3 \\\hline %99.6
 71--141  &  63.7$\pm$11.6 & 23.0$\pm$ 5.2 &  7.2$\pm$ 2.6 &  4.4$\pm$ 1.4 \\\hline %98.3
141--283  &  57.7$\pm$26.2 & 23.1$\pm$13.9 &  7.7$\pm$ 6.9 &  3.8$\pm$ 4.6 \\\hline\hline %92.3
\end{tabular}
\end{center}
\vskip-7mm
\end{table}

Table \ref{tab:sb1} shows in more detail the origin of $\gamma^{dir}$-candidates photons.
So, in Table \ref{tab:sb1}  the numbers in the  ``$\gamma-direct$'' column
correspond to the respective  numbers  of 
signal events in lines 1, 16 and 17  and column ``$S$'' of Table \ref{tab:sb4}
while the numbers in the ``$\gamma-brem$'' column
of Table  \ref{tab:sb1} correspond to the numbers
of events with the photons radiated from quarks
participating in the hard interactions. The total number of background events,
%(without events with electrons, last column ``$e^\pm$''), 
i.e. a sum over the numbers presented in columns 4 -- 8 in the same line, 
is shown in the column ``$B$'' of Table \ref{tab:sb4}.
The other lines of Table \ref{tab:sb1} for $\pth=40$ and
$~200 ~GeV/c~$ have the meaning analogous to that described above for $\pth=100 ~GeV/c$.

The last column of Table \ref{tab:sb1} shows the number of events with
$e^\pm$. In this paper we suppose the $100\%$ track finding efficiency 
\footnote{But, certainly, these electrons can be detected with the non-zero probability 
as a direct photon and their real contribution to the total background $B$ should be obtained
after account of the efficiency of charged tracks determination.}
for $e^\pm$ with $\Pt^e>40 ~GeV/c$.
%that can be detected with the non-zero probability as a direct photon.

%(as only small number of these events would be left).
The numbers in Tables \ref{tab:sb2} and \ref{tab:sb3}
accumulate in a compact form the information of Table \ref{tab:sb4} and \ref{tab:sb1}.
Thus, for example, the columns $S$ and $B$ of the middle lines for $\pth=100 ~GeV/c$  contain 
the numbers of the signal and background events taken at the level of line 16 (for Table
\ref{tab:sb2}) and line 17 (for Table \ref{tab:sb3}). 

From Table \ref{tab:sb2} it is seen that the ratio $S/B$ grows  while
$\Pt^{\tilde{\gamma}}$ increases from 3.9  at $\Pt^{\tilde{\gamma}}\geq 40 ~GeV/c$ to
48.4 at $\Pt^{\tilde{\gamma}}\geq 200 ~GeV/c$.
The jet isolation requirement (cut 17 from Table \ref{tab:sb0})
noticeably improves the situation at low $\Pt^{\tilde{\gamma}}$ (see Table \ref{tab:sb3}).
After application of this criterion $S/B$ increases
to 5.1 at $\Pt^{\tilde{\gamma}}\geq 40 ~GeV/c$ and to 24.46
at $\Pt^{\tilde{\gamma}}\geq 100 ~GeV/c$.
Remember the conclusion  that the sample of events
selected with our criteria has a tendency to contain more events with an isolated jet
as $\Pt^{\tilde{\gamma}}$ increases.

So, from  Tables \ref{tab:sb1} -- \ref{tab:sb3} we see that the cuts
listed in Table \ref{tab:sb0} (containing  moderate values of
$\Pt^{clust}_{CUT}$ and  $\Pt^{out}_{CUT}$) allow 
the major part of the background events to be suppressed.
The influence of wide variation of these two cuts on \\
(a) the number of selected events (for $L_{int}=3\,fb^{-1}$);\\
(b) the signal-to-background ratio $S/B$;\\
(c) the mean values of $(\Pt^{\tilde{\gamma}}\!-\!\Pt^{Jet})/\Pt^{\tilde{\gamma}}$ and
its  standard deviation value $\sigma (F)$\\
\\[-7mm]
\normalsize
\begin{table}[h]
\small
\begin{center}
\caption{Number of signal and background events remained after cuts.}
\vskip-2mm
\begin{tabular}{||c|c||c|c|c|c|c|c|c||}                  \hline \hline
\label{tab:sb1}
\hmm$\pth$\hmm& &$\gamma$ & $\gamma$ &\multicolumn{4}{c|}{  photons from the mesons}  &
\\\cline{5-8}
\Gvc& Cuts&\hmm direct\hmm &\hmm brem\hmm & $\;\;$ $\pi^0$ $\;\;$ &$\quad$ $\eta$ $\quad$ &
$\omega$ &  $K_S^0$ &\hmm $e^{\pm}$\hmm \\\hline \hline
    &Preselected&\hmm12394&\hmm 20952& 166821& 66533& 17464& 23942&\hmm 6684\hmm  \\\cline{2-9}
 40 &After cuts &\hmm 1718&\hmm 220&     146&     56&     2& 15&\hmm   10\hmm\\\cline{2-9}
    &+ jet isol.  &\hmm 1003&\hmm 102&    59&     26&     2&  7&   8\\\hline  \hline
    &Preselected&\hmm19359  &\hmm90022 &658981 &247644 &69210  & 85568 &\hmm47061\hmm\\\cline{2-9}
100 &After cuts&\hmm 1179 &\hmm34 &13 &4 &1  & 0 &\hmm 22\hmm \\\cline{2-9}   %out<10
    &+ jet isol. &\hmm 1125 &\hmm32 &9 &4 &1  & 0 &\hmm 21\hmm \\\hline \hline
    &Preselected&\hmm55839 &\hmm354602 &1334124 &393880 &141053 & 167605 &\hmm153410\hmm\\\cline{2-9}
200 &After cuts&\hmm 1838 &\hmm 27& 5 &5 &0  & 1 &\hmm 17\hmm\\\cline{2-9}
    &+ jet isol. &\hmm1831 &\hmm127& 5 &5 &0  & 1 &\hmm 17\hmm\\\hline \hline
\end{tabular}
\vskip0.2cm
\caption{Efficiencies and significance values in events without jet isolation cut.}
\vskip0.1cm
\begin{tabular}{||c||c|c|c|c|>{\columncolor[gray]{\coltab}}c|c||} \hline \hline
%\begin{tabular}{||c||c|c|c|c|c|c||} \hline \hline
\label{tab:sb2}
$\pth$ \Gvc& $S$ & $B$ & $Eff_S(\%)$  & $Eff_B(\%)$  & $S/B$& $S/\sqrt{B}$
\\\hline \hline
40  & 1718& 439 & 13.86 $\pm$ 0.36 & 0.149 $\pm$ 0.007&  3.9 & 82.0
 \\\hline
100 & 1179& 52 & 6.09 $\pm$ 0.18 & 0.005 $\pm$ 0.001& 22.7 & 163.5
 \\\hline  %out<10 GeV
200 & 1838& 38 & 3.29 $\pm$ 0.09 & 0.004 $\pm$ 0.001& 48.4 & 298.2 
\\\hline \hline
\end{tabular}
\vskip0.2cm
\caption{Efficiencies and significance values in events with jet isolation cut.}
\vskip0.1cm
\begin{tabular}{||c||c|c|c|c|>{\columncolor[gray]{\coltab}}c|c||}  \hline \hline
%\begin{tabular}{||c||c|c|c|c|c|c||}  \hline \hline
\label{tab:sb3}
$\pth$ \Gvc& ~~$S$~~ & ~~$B$~~ & $Eff_S(\%)$ & $Eff_B(\%)$  & $S/B$& $S/\sqrt{B}$
 \\\hline \hline
40  & 1003& 196 & 8.09 $\pm$ 0.27 & 0.066 $\pm$ 0.005&  5.1 & 71.6 \\\hline %out<10 GeV
100 & 1125&46 & 5.81 $\pm$ 0.18 & 0.004 $\pm$ 0.001& 24.5 & 165.9 \\\hline  %out<10 GeV
200 & 1831& 38 & 3.29 $\pm$ 0.09 & 0.004 $\pm$ 0.001& 48.4 & 298.2 %out<10 GeV
\\\hline \hline
\end{tabular}
\vskip-5mm
\end{center}
\end{table}

\noindent
is presented in Tables \ref{tab:b401} -- \ref{tab:b205} of Appendix.
Cuts (1) -- (10) of Table \ref{tab:sb0} of this section were applied to select
``direct photon candidate + 1 jet'' events for the tables of this Appendix.
The jets in these events as well as clusters were found by only one LUCELL jetfinder
(for the whole $\eta$ region $|\eta^{jet}|<5.0$).

Tables \ref{tab:b401} -- \ref{tab:b405} of Appendix 
correspond to the simulation with
$\pth=40 ~GeV/c$  and Tables  \ref{tab:b201} -- \ref{tab:b205} to that with
$\pth=200 ~GeV/c$. The  rows and  columns of Tables
\ref{tab:b401} -- \ref{tab:b205} illustrate the influence of
$\Pt^{clust}_{CUT}$ and $\Pt^{out}_{CUT}$ on the quantities
mentioned above (in the points (a), (b), (c)).

First of all, we see from Tables \ref{tab:b402} and \ref{tab:b202} of Appendix that
a noticeable reduction
of the background take place while moving along the diagonal from the right-hand bottom corner to the
left-hand upper one, i.e. with reinforcing $\Pt^{clust}_{CUT}$ and $\Pt^{out}_{CUT}$. So, we see that
for $\pth=40 ~GeV/c$ the ratio $S/B$ changes in the table cells along the diagonal
from $S/B=2.3$ (in the case of no limits on these two variables), to $S/B=3.9$ for the
cell with $\Pt^{clust}_{CUT}=10~ GeV/c$ and $\Pt^{out}_{CUT}=10 ~GeV/c$.
Analogously, for $\pth=200 ~GeV/c\,$ $S/B$ changes for the same table cells
from 13.6 to 48.4 (see the figures in Table \ref{tab:b202} of Appendix).
%for the same values of $\Pt^{clust}_{CUT}$ and $\Pt^{out}_{CUT}$.\\

The second observation. The restriction of $\Pt^{clust}_{CUT}$ and
$\Pt^{out}_{CUT}$ improves the calibration accuracy. Table
\ref{tab:b404} of Appendix  shows that the mean value of the fraction
$F\equiv (\Pt^{\tilde{\gamma}}\!-\!\Pt^{Jet})/\Pt^{\tilde{\gamma}}$
 decreases from 0.030 (the bottom right-hand corner) to 0.009
for $\Pt^{clust}_{CUT}=10~ GeV/c$ and $\Pt^{out}_{CUT}=10 ~GeV/c$.
Simultaneously, by this restriction one noticeably decreases 
% that include the standard deviation values),
(about a factor of two: from 0.163 to 0.085 for $\pth=40 ~GeV/c$, for instance)
the width of the gaussian $\sigma (F)$ (see Tables \ref{tab:b405} and
\ref{tab:b205} of Appendix).

%\noindent
The explanation is simple. The balance~ equation (\ref{eq:sc_bal}) contains 2 terms on the right-hand
side ($1-cos\dphi$) and $\Db/\Pt^{\gamma}$.
The first one is negligibly small and tends to decrease with growing 
$\Pt^{\tilde{\gamma}}$ (see \cite{BKS_P1} for details). So, we see that
the main source of the disbalance in  equation (\ref{eq:sc_bal}) is the term $\Db/\Pt^{\tilde{\gamma}}$.
This term can be decreased by decreasing
$\Pt$ activity beyond the jet.% which leads to improvement of the calibration accuracy.

Thus, we can conclude that application of two criteria introduced
in Section 2, i.e. $\Pt^{clust}_{CUT}$ and $\Pt^{out}_{CUT}$,
results in two important consequences: significant background reduction
and essential improvement of the calibration accuracy.

The numbers of events (for $L_{int}= 3 ~fb^{-1}$)
for different $\Pt^{clust}_{CUT}$ and $\Pt^{out}_{CUT}$
are given in the cells of Tables \ref{tab:b401} and
\ref{tab:b201} of Appendix. One can see that even with such strict
$\Pt^{clust}_{CUT}$ and $\Pt^{out}_{CUT}$ values as $10 ~GeV/c$ for both, for example,
we would have a sufficient number of events
(about 3 600 000  for $\Pt^{\tilde{\gamma}}\geq40 ~GeV/c$,
and 4 000 $\Pt^{\tilde{\gamma}}\geq200 ~GeV/c$)
with low background contamination ($S/B=3.9$ and $48.4$ for 
$\Pt^{\tilde{\gamma}}\geq40 ~GeV/c$ and $\Pt^{\tilde{\gamma}}\geq200 ~GeV/c$ respectively)
and a good accuracy of the absolute jet energy scale setting.

Let us mention that all these PYTHIA results can serve as
preliminary ones and only full GEANT simulation would allow one to
come to a final conclusion. % to draw an inference from smth. 

To conclude this section we would like to stress that, as is seen from
Table~\ref{tab:sb1}, the ``$\gamma-brem$'' background
defines a dominant part of the total background. Its contribution
%remained after cuts 1 -- 17 of Table~\ref{tab:sb0}
is about the same (see Table~\ref{tab:sb1}) as
the combined background contribution from neutral meson decays.
%Thus, one can see from Table 17 that $\pi^0$ contribution being about a half of
%``$\gamma-brem$'' background at $\pth>40~GeV/c$ becomes one order less than
%``$\gamma-brem$'' background at $\pth>200~GeV/c$. 
We would like to emphasize here
that this is a strong prediction of PYTHIA generator which has to be compared with
predictions of other generator like HERWIG, for example.

Secondly, we would like to underline also that as it is seen from Table 14, 17 the photon
isolation and selection cuts 1--5, usually used in the study of inclusive photon production,
 increase the $S/B$ ratio up to
1.93 only while the other cuts 6--17, that select events with a clear \gpj topology and
limited $\Pt$ activity beyond a chosen single jet, lead to a significant improvement of
$S/B$ by about one order of magnitude to $24.46$.

The numbers in the tables of Appendix were obtained with inclusion of the contribution
from the background events. The tables show that they account does not spoil the \ptgj
balance. The estimation of the number of these background events would be important
for the gluon distribution determination (see Section 4).

\normalsize

%\newpage
%=======================================================================
%
\section{\gpj event rate estimation for gluon distribution determination at the LHC.}
%        Chapter 10.
%=======================================================================

As many of theoretical predictions for production of new particles
(Higgs, SUSY) at the LHC
are based on model estimations of the gluon density behavior at
low $x$ and high $Q^2$, measurement of the proton gluon density
for this kinematic region directly in LHC experiments
would be obviously useful. One of the promising
channels for this measurement, as was shown in \cite{Au1},
is a high $\Pt$ direct photon production $pp(\bar{p})\rightarrow \gamma^{dir} + X$.
The region of high $\Pt$, reached by UA1, UA2, CDF and
D0 extends up to $\Pt \approx 60~ GeV/c$ and recently up to $\Pt=105~GeV/c$ \cite{D0_2}. 
These data together with the later ones and recent
E706 and UA6 results give an opportunity for tuning the form of gluon 
distribution. 

Here for the same aim we shall consider %(see also \cite{MD1}) 
the process $pp\rightarrow \gamma^{dir}+ 1 Jet + X$
defined in the leading order by two QCD subprocesses (\ref{eq:1a}) and (\ref{eq:1b}).

%\noindent
The ``$\gamma^{dir}+1 \,Jet$'' final state is
more preferable than inclusive photon production $\gamma + X$ from
the viewpoint of extraction of information on gluon distribution.
Indeed, in the case of inclusive direct photon production the
cross section is given as an integral over partons distribution
functions $f_a(x_a,Q^2)$ (a = quark or gluon), while in the case of
$pp\to\gamma^{dir}+1\,Jet+X$ for $\Pt^{Jet}\,
\geq \,30\, GeV/c$ (i.e. in the region where ``$k_t$'' smearing
effects are not important) the cross section is
expressed directly in terms of these distributions (see, for example,
\cite{Owe}) \\[-15pt]
\begin{eqnarray}
\frac{d\sigma}{d\eta_1d\eta_2d\Pt^2} = \sum\limits_{a,b}\,x_a\,f_a(x_a,Q^2)\,
x_b\,f_b(x_b,Q^2)\frac{d\sigma}{d\hat{t}}(a\,b\rightarrow 3\,4)
\label{gl:4}
\end{eqnarray}
\vskip-3mm
\noindent
where \\[-9mm]
\begin{eqnarray}
x_{a,b} \,=\,\Pt/\sqrt{s}\cdot \,(exp(\pm \eta_{1})\,+\,exp(\pm \eta_{2})).
\label{eq:x_def}
\end{eqnarray}
We also used the following designations above:
$\eta_1=\eta^\gamma$, $\eta_2=\eta^{Jet}$; ~$\Pt=\Pt^\gamma$;~ a,b = $q, \bar{q},g$; 
3,4 = $q,\bar{q},g,\gamma$.
Formula (\ref{gl:4}) and the knowledge of the results of independent measurements of
$q, \,\bar{q}$ distributions %\cite{MD1}
allow the gluon $f_g(x,Q^2)$ distribution
to be determined after account of selection efficiencies of $\gamma^{dir}$ candidates 
and the contribution of background, left after the used selection cuts (1--13 of Table \ref{tab:sb0}),
as it was discussed in Section 3 keeping in mind this task.

In the previous sections a lot of details connected with the
structure and topology of these events and the objects appearing
in them were discussed. Now with this information in mind we are
in position to discuss application of the \gpj
event samples selected with the proposed cuts to estimate
rates of gluon-based  subprocess (\ref{eq:1a}).

In Table~\ref{tab:q/g-2} we present the $Q^2 (\equiv(\Ptg)^2)$ and $x$ (defined according to (\ref{eq:x_def}))
distribution of the number of events that are caused by the $q~ g\rrr \gamma +q$ subprocess, 
and passed cuts (\ref{eq:sc1}) -- (\ref{eq:sc8}) of Section 2 ($\Pt^{out}$ was not limited):\\[-7mm]
\begin{eqnarray}
\Ptg>40~ GeV/c, |\eta^{\gamma}|<2.5, \Pt^{Jet}>30~ GeV/c, |\eta^{Jet}|<5.0, \Pt^{hadr}>5~ GeV/c,
\label{l1}
\nonumber
\end{eqnarray}
~\\[-15mm]
\begin{eqnarray}
\Pt^{isol}_{CUT}=5\;GeV/c, \;
{\epsilon}^{\gamma}_{CUT}=7\%, \;
\dphi<15^{\circ}, \; 
\Pt^{clust}_{CUT}=5\;GeV/c. \;
%\Pt^{out}_{CUT}=20\;GeV/c.
\label{l2}
\end{eqnarray}

\begin{table}[h]
\vskip-3mm
\begin{center}
\vskip-0.2cm
\small
\caption{Number of~ $g\,q\to \gamma^{dir} \,+\,q$~
events at different $Q^2$ and $x$ values for $L_{int}=20~fb^{-1}$.}
\label{tab:q/g-2}
\vskip0.2cm
\begin{tabular}{|lc|c|c|c|c|c|c|}                  \hline
 & $Q^2$ &\multicolumn{4}{c|}{ \hspace{-0.9cm} $x$ values of a parton} & All $x$   \\\cline{3-7}
 & $(GeV/c)^2$ & $10^{-4}$--$10^{-3}$ & $10^{-3}$--$10^{-2}$ &$10^{-2}$--
$10^{-1}$ & $10^{-1}$--$10^{0}$ & $10^{-4}$--$10^{0}$     \\\hline
&\hmm\hmm 1600-2500\hmm  & 735.7 &2319.2 &2229.0 & 236.9 &5521.0 \\\hline
&\hmm\hmm 2500-5000\hmm  & 301.6 &1323.3 &1402.7 & 207.4 &3235.1 \\\hline
&\hmm\hmm 5000-10000\hmm &  33.7 & 361.3 & 401.0 &  97.7 & 893.8 \\\hline
&\hmm\hmm 10000-20000\hmm&   1.5 &  80.8 &  99.4 &  38.0 & 219.9 \\\hline
&\hmm\hmm 20000-40000\hmm&     0 &  15.6 &  24.4 &  12.4 &  52.5 \\\hline
&\hmm\hmm 40000-80000\hmm&     0 &   2.1 &   4.2 &   2.5 &   8.8 \\\hline
\end{tabular}
\end{center}
\end{table}

\begin{table}[h]
\begin{center}
\vskip-9mm
\small
\caption{Number of~ $g\,c\to \gamma^{dir} \,+\,q$~
events at different $Q^2$ and $x$ values for $L_{int}=20~fb^{-1}$.}
\label{tab:q/g-3}
\vskip.2cm
\begin{tabular}{|lc|c|c|c|c|c|c|c|} \hline
& $Q^2$ &\multicolumn{4}{c|}{ \hspace{-1.2cm} $x$ values for $c$-quark} & All $x$ \\\cline{3-7}
& $(GeV/c)^2$ & $10^{-4}$--$10^{-3}$ & $10^{-3}$--$10^{-2}$ &$10^{-2}$--
$10^{-1}$ & $10^{-1}$--$10^{0}$ & $10^{-4}$--$10^{0}$  \\\hline
&\hmm\hmm 1600-2500\hmm   &109.4 & 360.5 & 329.6 &  34.7 & 834.4 \\\hline
&\hmm\hmm 2500-5000\hmm   & 35.1 & 189.7 & 202.7 &  25.4 & 453.2 \\\hline
&\hmm\hmm 5000-10000\hmm  &  3.9 &  51.5 &  58.6 &  12.1 & 126.3 \\\hline
&\hmm\hmm 10000-20000\hmm &  0.1 &   9.0 &  12.4 &   3.4 &  25.0 \\\hline
&\hmm\hmm 20000-40000\hmm &    0 &   1.4 &   3.2 &   1.0 &   5.6 \\\hline
&\hmm\hmm 40000-80000\hmm &    0 &   0.1 &   0.4 &   0.1 &   0.7 \\\hline
\end{tabular}
\end{center}
\vskip-4mm
\end{table}

\normalsize
The analogous information for events with the charmed quarks in the initial state
$g\,c\to \gamma^{dir} \,+\,c~$ is presented in Table~\ref{tab:q/g-3}. 
The simulation of the process $g\,b\to \gamma^{dir}\,+\,b$ $\;$ shows that the rates
for the $b$-quark are 8 -- 10 times smaller than for
the $c$-quark.

Fig.~3 shows in the widely used $(x,Q^2)$
kinematic plot what area can be covered by studying the process $q~ g\rrr \gamma +q$.
The number of events in this area 
\begin{flushright}
\vskip-4.2mm
\parbox[r]{.49\linewidth}
{ \vspace*{0.1cm}
 \hspace{0mm} 
is given in Table~\ref{tab:q/g-2}.
From this figure and Table~\ref{tab:q/g-2}
 it becomes clear that even at integrated luminosity
$L_{int}=20~fb^{-1}$ it would be possible to study the gluon distribution with good statistics 
of \gpj events in the region of small $x$
at $Q^2$ about 2--3  orders of magnitude higher than now reached at HERA.
It is worth emphasising that extension of the experimentally reachable
region at the LHC to the region of lower $Q^2$ overlapping with the area
covered by HERA would also be of great interest.}
\end{flushright}
%\begin{flushleft}
\begin{figure}[h]
   \vskip-79.1mm
   \hspace{-1mm} \includegraphics[width=.51\linewidth,height=76mm,angle=0]{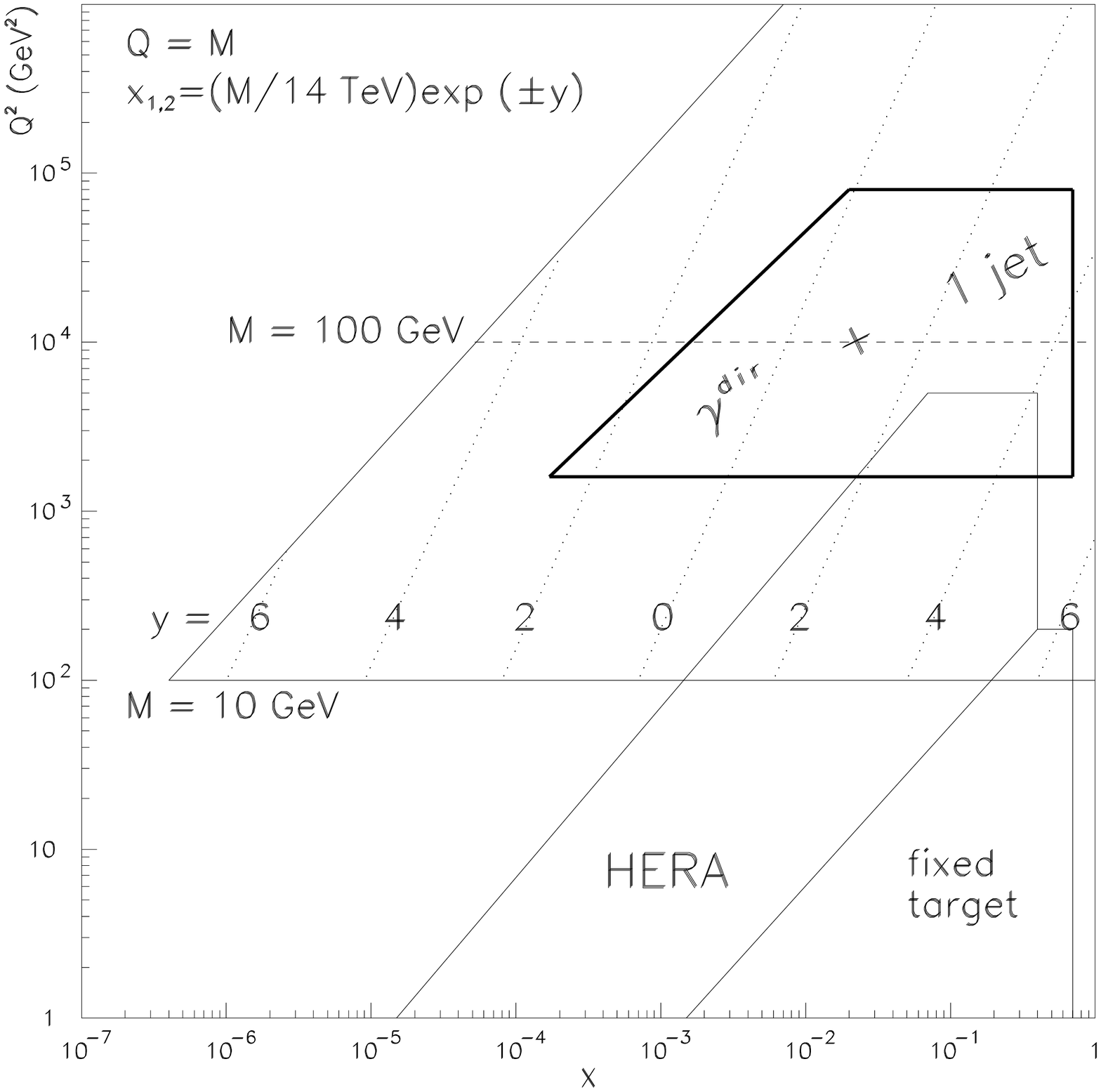}
\label{fig:q/g}
\end{figure}
%\end{flushleft}
%\hspace*{90mm}
{\vskip-0.3cm
\hspace*{.5cm} {Figure~3: \footnotesize {The  $(x,Q^2)$ kinematic region for
$pp\to \gamma+Jet$ process.}}
}

%\vskip-3mm

\setcounter{table}{0}

 %                                 40
\begin{table}[h]
\vskip-0mm
%---------------------------------------------------------------------
% 					                             |
\hspace*{0mm} {\bf \large Appendix}\\[-8pt]                                             %|
% 3.3					                             |
%---------------------------------------------------------------------
%%%%%%%%%%%                40 GeV
\def\baselinestretch{0.99}
%\baselineskip-20pt
%\begin{table}[htbp]
\small
%\vskip-20mm
\begin{center}
\vskip-7mm
\large{ ~$~~\pth= 40~GeV/c$}\\[5pt]
\large{ $\Pt^{isol}<2 ~GeV/c,~~ \epsilon^{\tilde{\gamma}}<5\%,~~ \Delta\phi=15^\circ$}\\[15pt]
\normalsize
\vskip-0.1cm
\small
\caption{Number of events per $L_{int}=3~fb^{-1}$}
\label{tab:b401}
%\vskip0.0cm
\begin{tabular}{|c||c|c|c|c|c|c|} \hline
%\hline
$\Pt^{clust}_{\,cut}$ &\multicolumn{6}{c|}{$\Pt^{\,out}_{\,cut}~(GeV/c)$}
\\\cline{2-7}
\Gvc  &\han 5\han&\han 10\han&\han 15\han&\han 20\han&\han 30\han&
 1000  \\\hline \hline
    5&   634000&  1064000&  1108000&  1110000&  1110000&  1110000\\\hline
   10&  1681000&  3625000&  4382000&  4578000&  4616000&  4618000\\\hline
   15&  1939000&  4548000&  6051000&  6641000&  6813000&  6822000\\\hline
   20&  2017000&  4893000&  6756000&  7674000&  8081000&  8121000\\\hline
   30&  2090000&  5140000&  7258000&  8456000&  9317000&  9581000\\\hline
\end{tabular}
\vskip0.2cm
\small
\caption{ $S/B$}
\label{tab:b402}
\begin{tabular}{|c||c|c|c|c|c|c|} \hline
%\hline
$\Pt^{clust}_{\,cut}$ &\multicolumn{6}{c|}{$\Pt^{\,out}_{\,cut}~(GeV/c)$}
\\\cline{2-7}
\Gvc  &\has 5\has&\has 10\has&\has 15\has&\has 20\has&\has 30\has&\hass
 1000 \hass \\\hline \hline
    5&   5.6$\pm$ 1.1&   5.0$\pm$ 0.7&   4.8$\pm$ 0.7&   4.8$\pm$ 0.7&   4.8$\pm$ 0.7&   4.8$\pm$ 0.7\\\hline
   10&   4.2$\pm$ 0.5&   3.9$\pm$ 0.3&   3.6$\pm$ 0.2&   3.5$\pm$ 0.2&   3.5$\pm$ 0.2&   3.5$\pm$ 0.2\\\hline
   15&   3.7$\pm$ 0.4&   3.4$\pm$ 0.2&   3.2$\pm$ 0.2&   3.1$\pm$ 0.2&   3.0$\pm$ 0.2&   3.0$\pm$ 0.2\\\hline
   20&   3.7$\pm$ 0.4&   3.2$\pm$ 0.2&   2.9$\pm$ 0.2&   2.8$\pm$ 0.1&   2.7$\pm$ 0.1&   2.7$\pm$ 0.1\\\hline
   30&   3.5$\pm$ 0.3&   2.9$\pm$ 0.2&   2.6$\pm$ 0.1&   2.5$\pm$ 0.1&   2.3$\pm$ 0.1&   2.3$\pm$ 0.1\\\hline          
\end{tabular}
\vskip0.2cm
\small
\caption{$\la F\ra,~F= \Fptgj$ }
\label{tab:b404}
\begin{tabular}{|c||c|c|c|c|c|c|} \hline
%\hline\
$\Pt^{clust}_{\,cut}$ &\multicolumn{6}{c|}{$\qquad \qquad\Pt^{\,out}_{\,cut}~(GeV/c)\qquad \qquad$}
\\\cline{2-7}
\Gvc&\had 5\had &\had 10\had&\had 15\had&\had 20\had&\had 30\had& \hass 1000 \hass\\\hline\hline
   5&    0.008&    0.008&    0.008&    0.008&    0.008&    0.008\\\hline
   10&    0.003&    0.009&    0.013&    0.013&    0.013&    0.013\\\hline
   15&    0.005&    0.011&    0.018&    0.019&    0.022&    0.022\\\hline
   20&    0.006&    0.012&    0.020&    0.023&    0.026&    0.027\\\hline
   30&    0.005&    0.012&    0.021&    0.024&    0.029&    0.030\\\hline  
\end{tabular}
\vskip0.2cm
\small
\caption{$\sigma(F),~F= \Fptgj$ }
\label{tab:b405}
\begin{tabular}{|c||c|c|c|c|c|c|} \hline
%\hline\
$\Pt^{clust}_{\,cut}$ &\multicolumn{6}{c|}{$\qquad \qquad\Pt^{\,out}_{\,cut}~(GeV/c)\qquad \qquad$}
\\\cline{2-7}
\Gvc&\had 5\had &\had 10\had&\had 15\had&\had 20\had&\had 30\had& \hass 1000 \hass\\\hline\hline
    5&    0.063&    0.075&    0.079&    0.079&    0.079&    0.079\\\hline
   10&    0.068&    0.085&    0.097&    0.102&    0.104&    0.104\\\hline
   15&    0.070&    0.090&    0.109&    0.123&    0.129&    0.130\\\hline
   20&    0.070&    0.092&    0.113&    0.133&    0.145&    0.147\\\hline
   30&    0.071&    0.093&    0.117&    0.140&    0.159&    0.163\\\hline  
\end{tabular}
\end{center}
\end{table}

%%%%%%%%%%%                200 GeV
\def\baselinestretch{0.95}
\begin{table}[htbp]
\small
\begin{center}
\vskip-1mm
\large{ ~$~~\pth= 200~GeV/c$}\\[5pt]
\large{ $\Pt^{isol}<2 ~GeV/c,~~ \epsilon^{\tilde{\gamma}}<5\%,~~ \Delta\phi=15^\circ$}\\[15pt]
\normalsize
\small
\caption{Number of events per $L_{int}=3~fb^{-1}$}
\label{tab:b201}
\begin{tabular}{|c||c|c|c|c|c|c|} \hline
%\hline
$\Pt^{clust}_{\,cut}$ &\multicolumn{6}{c|}{$\Pt^{\,out}_{\,cut}~(GeV/c)$}
\\\cline{2-7}
\Gvc  &\hcn 5\hcn&\hcn 10\hcn&\hcn 15&\hcn 20\hcn&\hcn 30\hcn&\hcn 1000\hcn \\\hline\hline
    5&      620&     1220&     1330&     1360&     1360&     1380\\\hline
   10&     1660&     4100&     5220&     5700&     5820&     5840\\\hline
   15&     2080&     5420&     7880&     9310&    10160&    10290\\\hline
   20&     2230&     5960&     9020&    11240&    13230&    13840\\\hline
   30&     2310&     6290&     9770&    12590&    16570&    19510\\\hline                 
\end{tabular}
\vskip0.2cm
\small
\caption{$S/B$}
\label{tab:b202}
\begin{tabular}{|c||c|c|c|c|c|c|} \hline
% \hline
$\Pt^{clust}_{\,cut}$ &\multicolumn{6}{c|}{$\Pt^{\,out}_{\,cut}~(GeV/c)$}
\\\cline{2-7}
\Gvc  & 5& 10& 15& 20& 30& 1000  \\\hline \hline
    5& 179o 165& 114$\pm$61& 102$\pm$49& 104$\pm$50& 104$\pm$50& 104$\pm$50\\\hline
   10&  48.9$\pm$12.4&  48.4$\pm$ 8.6&  47.2$\pm$ 7.6&  45.7$\pm$ 6.0&  45.5$\pm$ 6.1&  45.5$\pm$ 6.1\\\hline
   15&  42.1$\pm$11.2&  42.8$\pm$ 7.1&  39.9$\pm$ 5.3&  31.5$\pm$ 3.5&  28.4$\pm$ 2.9&  28.3$\pm$ 2.9\\\hline
   20&  31.2$\pm$ 7.0&  36.1$\pm$ 5.3&  29.7$\pm$ 3.3&  24.7$\pm$ 2.3&  20.7$\pm$ 1.6&  19.4$\pm$ 1.5\\\hline
   30&  30.2$\pm$ 6.6&  28.6$\pm$ 3.7&  23.2$\pm$ 2.2&  19.3$\pm$ 1.5&  15.8$\pm$ 1.0&  13.6$\pm$ 0.8\\\hline
\end{tabular}
\vskip0.2cm
\small
\caption{$\la F\ra, ~F=\Fptgj$}
\label{tab:b204}
\begin{tabular}{|c||c|c|c|c|c|c|} \hline
\\\cline{2-7}
\Gvc  &\hcd 5\hcd&\hcd 10\hcd&\hcd 15\hcd&\hcd 20\hcd&\hcd 30\hcd&
\hcd 1000 \hcd \\\hline \hline
    5&    0.003&    0.002&    0.003&    0.003&    0.003&    0.004\\\hline
   10&    0.001&    0.003&    0.004&    0.005&    0.005&    0.005\\\hline
   15&    0.001&    0.003&    0.005&    0.007&    0.007&    0.008\\\hline
   20&    0.001&    0.003&    0.005&    0.007&    0.008&    0.009\\\hline
   30&    0.001&    0.003&    0.005&    0.007&    0.009&    0.014\\\hline
\end{tabular}
\vskip0.2cm
\small
\caption{$\sigma(F),~F= \Fptgj$ }
\label{tab:b205}
\begin{tabular}{|c||c|c|c|c|c|c|} \hline
%\hline\
$\Pt^{clust}_{\,cut}$ &\multicolumn{6}{c|}{$\qquad \qquad\Pt^{\,out}_{\,cut}~(GeV/c)\qquad \qquad$}
\\\cline{2-7}
\Gvc  &\hcd 5\hcd&\hcd 10\hcd&\hcd 15\hcd&\hcd 20\hcd&\hcd 30\hcd&
\hcd 1000 \hcd \\\hline
%\hline
    5&    0.014&    0.017&    0.019&    0.020&    0.022&    0.024\\\hline
   10&    0.015&    0.019&    0.023&    0.025&    0.027&    0.027\\\hline
   15&    0.015&    0.020&    0.025&    0.029&    0.033&    0.035\\\hline
   20&    0.015&    0.021&    0.026&    0.031&    0.038&    0.042\\\hline
   30&    0.015&    0.021&    0.027&    0.033&    0.043&    0.054\\\hline   
\end{tabular}
\end{center}
\end{table}


\normalsize

\begin{thebibliography}{99}
%==================================================
% main refs:      O S N O V N Y E    S S Y L K I  zdes' ----
%==================================================         |
\bibitem{1}
N.B.~Skachkov, V.F.~Konoplyanikov D.V.~Bandourin,
``Photon -- jet events for calibration of HCAL''. Second Annual RDMS CMS Collaboration Meeting.
CMS Document 1996--213. CERN, December 16-17, 1996, p.7-23.
\bibitem{2}
N.B.~Skachkov, V.F.~Konoplyanikov D.V.~Bandourin,
``$\gamma$-direct + 1 jet events for HCAL calibration''.
Third Annual RDMS CMS Collaboration Meeting. CMS Document
1997--168. CERN, December 16-17, 1997, p.139-153.
\bibitem{3}
 D.V. Bandurin, V.F. Konoplyanikov, N.B. Skachkov,
"Jet energy scale setting with "gamma+jet" events for a hadronic calorimeter of CMS."
Fifth Annual RDMS CMS Collaboration Meeting. CMS Document 2000-058. Conference. 
"Physics Program with the CMS Detector",
ITEP, Moscow, Russia. November 22-24, 2000. p. 422-427.
\bibitem{BKS_P1}
D.V.~Bandourin, V.F.~Konoplyanikov, N.B.~Skachkov.
``Jet energy scale setting with \gpj events at LHC energies.''
JINR Preprints E2-2000-251 -- E2-2000-255, JINR, Dubna, hep-ex/0011012,
hep-ex/0011013, hep-ex/0011014, hep-ex/0011017, hep-ex/0011084.
\bibitem{BKS_GLU}
 D.V.~Bandourin, V.F.~Konoplyanikov, N.B.~Skachkov,
``\gpj events rate estimation for gluon distribution determination at LHC'',
Part.Nucl.Lett.103:34-43,2000, hep-ex/0011015.
\bibitem{CMS_EC}
CMS Electromagnetic Calorimeter Project, Technical Design Report,
CERN/LHCC 97--33, CMS TDR 4, CERN, 1997.
\bibitem{PYT}
T.~Sjostrand, Comp.Phys.Comm. {\bf 82} (1994)74.
\bibitem{Fri}
 S.~Frixione, Phys.Lett. {\bf B429} (1998)369.
\bibitem{Cat}
S.~Catani, M.~Fontannaz and E.~Pilon, Phys.Rev. {\bf D58} (1998)094025.
%========================
% gluon refs:
%========================
\bibitem{Au1} %1
P.~Aurenche {\it et al.}
Proc. of "ECFA LHC Workshop", Aachen, Germany, 4-9 Oktob. 1990,
edited by G.~Jarlskog and D.~Rein (CERN-Report No 90-10; Geneva, Switzerland
1990), Vol. {\bf II}
\bibitem{D0_2} %5
 D0 Collaboration, B.~Abbott {\it et al.} , Phys.Rev.Lett. {\bf 84} 2786-2791,2000.
\bibitem{Owe} %7
 J.F.~Owens, Rev.Mod.Phys. {\bf 59} (1987)465.

\end{thebibliography}
\end{document}